\documentclass[twocolumn,showpacs,preprintnumbers,amsmath,amssymb]{revtex4}
\usepackage{dcolumn}
\usepackage{bm}
\usepackage{amsmath}
\usepackage{amssymb}
\usepackage{latexsym}
\usepackage[applemac]{inputenc}
\usepackage[dvips]{graphicx}

\begin{document}


\title{To grate a liquid into tiny droplets by its impact on a hydrophobic micro-grid}

\author{P. Brunet}
\email{philippe.brunet@univ-lille1.fr}
\author{F. Lapierre}
\author{F. Zoueshtiagh}
\author{V. Thomy}
\author{A. Merlen}

%

\affiliation{Institut d'Electronique de Micro\'electronique et de Nanotechnologies, UMR CNRS 8520, Cit\'e Scientifique, Avenue Poincar\'e, BP. 60069, 59652 Villeneuve d'Ascq, France%
}%

\date{\today}

\begin{abstract}

We report on experiments of drop impacting a hydrophobic micro-grid, of typical spacing a few tens of $\mu$m. Above a threshold in impact speed, liquid emerges to the other side, forming micro-droplets of size about that of the grid holes. We propose a method to produce either a mono-disperse spray or a single tiny droplet of volume as small as a few picoliters corresponding to a volume division of the liquid drop by a factor of up to 10$^5$. We also discuss the discrepancy of the measured thresholds with that predicted by a balance between inertia and capillarity.

\end{abstract}


\maketitle

\begin{figure}[tb]
\begin{center}
\includegraphics[scale=0.19]{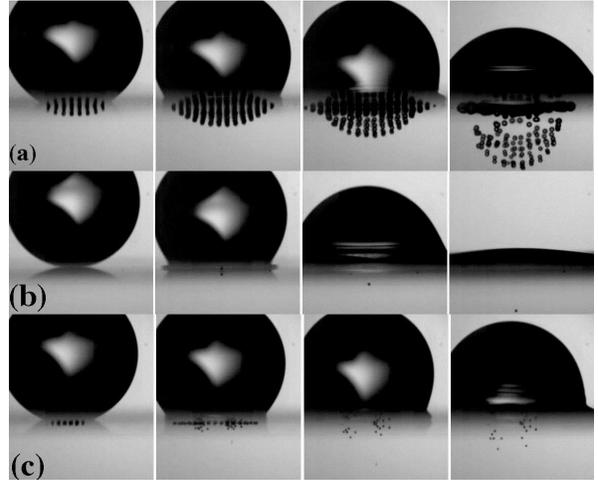}
\caption{Successive shots showing the impact of a water drop for different grids and impact velocity $U$. (a) $U$ = 1.60 m/s, Grid $\#$2 and the subsequent 'rain' of droplets emerging below the grid. (b) A sequence at threshold: a single tiny droplet emerges down the grid. $U$ = 1.94 m/s, Grid $\#$1 inverse. (c) A double-peaked distribution of emerging droplets. $U$ = 1.27 m/s, Grid $\#$1.}
\label{fig:mediumgrid_seq}
\end{center}
\end{figure}

For discrete microfluidics, it is often desired to produce smaller and smaller quantities of liquid in a controlled fashion. Usual methods are the generation of secondary droplets by drop impact \cite{Yarin06,Nonlinearity}, the breakup-up of shear-flow-mediated ligaments \cite{Marmottant04}, the appliance of intense electric fields to conductive liquids \cite{Ganancalvo,Basaran} or a high-frequency sound wave generated by a piezoelectric at the outlet of a nozzle \cite{Schneider64}. Here, we propose an alternative method: the impact of drops on a hydrophobic grid of micron-sized holes (diameter $d$). Figure \ref{fig:mediumgrid_seq} shows three sequences of such impacts: a few tens of $\mu$s after the drop impacted the grid, the liquid previously lying at the base of the drop is grated over and turned into small - quite monodisperse - droplets with different spatial distribution. We show that, despite the violent character of impact leading to complex flows, the emerging volume depends on the impact speed in a reproducible way. Understanding the basic mechanisms of how the liquid passes through, or is retained by, the grid has multiple applications in processes such as aerosols or filters. Generally, this simple geometry helps  to understand how liquid is captured by surface tension forces, during its contact on a porous solid of more complex shape. 

Previous studies on the impact of drops onto holes were carried out by Lorenceau and Qu\'er\'e \cite{Lorenceau03}. They used single-hole sieve of size ranging from 100 $\mu$m to 1.7 mm, smaller than the capillary length $l_c = \sqrt{\sigma / \rho g}$. The authors proposed that the threshold for protruding liquid results in a balance between initial liquid inertia and capillarity. Hence, the natural control parameter of the experiment is the Weber number: $\text{We} = \rho U^2 d / (2 \sigma)$, built with the impact velocity $U$, the typical size of the hole $d$, the surface tension $\sigma$ and density $\rho$ of the liquid. Due to the low-viscosity liquid we used (water), the Reynolds number is between 10 and 200: retention forces opposing liquid entry are mostly due to capillarity \cite{Lorenceau03}. However, we evidence that the impact on a network of holes leads to collective effects which contrasts from the single hole case. In short, the threshold for droplet detachment below the grid is much smaller than for a single hole. Also, the distribution of emerging droplets can show peculiar shapes, either single- (Fig.~\ref{fig:mediumgrid_seq}-(a)) or double-peaked (Fig.~\ref{fig:mediumgrid_seq}-(c)), suggesting a non-trivial pressure profile produced by the impact itself. In this letter, we show the results of drop impalement through various grids, focussing on the regimes where 1 to about 100 small droplets emerge. 

We use a dripping faucet that releases a drop of water (viscosity $\nu$ = 1 cSt, $\sigma$ = 0.072 N/m, $\rho$= 1 g/cm$^3$) from a submillimetric nozzle. The diameter of the impacting drop $D$ is determined by the capillary length $l_c$, and is reproducible at $D$ = 2.18 $\pm$ 0.05 mm. The height of fall $h$ prescribes the velocity at impact, $U = (2 g h)^{1/2}$, that can be up to 3 m/s. Using backlighting with a high-speed camera at a rate of 8600 frames/s, with a resolution of 288$\times$480 pixels, the shape of the interface during the spreading and bouncing processes can be determined. The magnification allows for an accuracy of about 8 $\mu$m per pixel. 

\begin{figure}[tb]
\begin{center}
\includegraphics[scale=0.20]{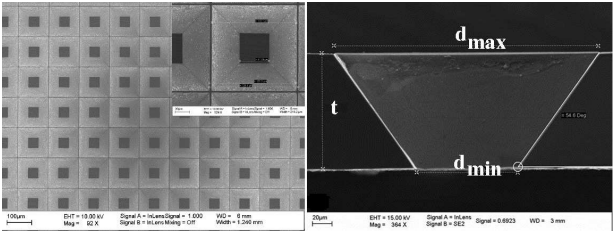}\\
\caption{SEM pictures (top and cross views) of a hydrophobic grid. The trapezoidal holes have $d_{min}$ and $d_{max}$ respectively as smaller and larger aperture.}
\label{fig:grid_process}
\end{center}
\end{figure}

The grids were fabricated in a clean room using the following process: a 200 nm layer of silicon nitride Si$_x$N$_y$, deposited with a low-pressure chemical vapor deposition (LPVCD) technique, coats both sides of a 380 $\mu$m silicon substrate in a $<$100$>$ plan oriented. A mask made of AZ1518 resin is used to design the grid by optical lithography. Then a CHF$_3$/CF$_4$ plasma creates grooves at locations where there is no resin, and removes the nitride layer at the back face. The substrate is rinsed with acetone and ethanol, and the resin is removed. The back side of the substrate, kept in a KOH bath during about 5 hours, is etched to a 100 $\mu$m layer substrate. Then, the whole substrate is thinned to 50 $\mu$m in the same bath. The etching following the $<$111$>$ oriented plan, is anisotropic and produces holes with trapezoidal shape. Finally, a hydrophobic coating on the overall surface of the grid is achieved using $C_4F_8$ plasma deposition. A typical resulting grid is shown in Fig.~\ref{fig:grid_process}. Grid sizes are summarised in Table \ref{tab:grids}. The interplay between the texture and the hydrophobic coating makes the grid highly water repellent: a drop sits on the grids with a spherical shape, rolling on with almost no friction. Despite their thickness, the grids were stiff enough not to be bent significantly during impact. 

\begin{table}[tb]
\caption{Dimensions of the different grids. $b$ is the space between holes, $a$ is the diameter of the emitted droplets.}
\begin{center}
\begin{ruledtabular}
\begin{tabular}{cccccc}
Grid  & Thickness & $d_{min}$  & $d_{max}$ & $b$ & $a$   \\
number & $t$ ($\mu$m)  & ($\mu$m) & ($\mu$m) & ($\mu$m) & ($\mu$m) \\
\space
\hline
1  &  49 &  22 &  93 & 117 & 36.5   \\
2  & 47 & 49 &  117 & 120 & 97.5  \\
3 &  38.5  &  38 &  91 & 121 & 69  \\
4 &  56.5  &  35 &  117 & 118 & 65  \\
5  &  56 &  10 &  93 & 119 & 16.5 \\
\end{tabular}
\end{ruledtabular}
\end{center}
\label{tab:grids}
\end{table}

The trapezoidal shape of the holes allows for testing the grids in both direct - where the drop impacts the widest apertures of the holes ($d_{max}$) and inverse configurations. Careful observations reveal that within the accuracy allowed by our visualisations, the detached droplets are all about the same size, hence constituting a monodisperse spray. Also, they are of same size for both direct and inverse impact configurations and do not depend on the Weber number. However, their size is strongly related to the smaller dimension of the trapezoidal holes $d_{min}$. For instance, we obtained the tiniest droplets with grid $\#$5 which gave a diameter of 16 $\mu$m corresponding to a volume of about 2 pl. The mechanism invoked for the droplets formation is the following: the liquid entering the holes, emerges at the other side in the form of tiny jets of diameter equal to about $d_{min}$. These jets are subject to the Rayleigh-Plateau (RP) instability that pinches them off into droplets \cite{Jens_rmp97}. For such tiny jets, the pinch-off is so fast that the liquid cylinder breaks-up before developing. The volume of a single droplet is equal to that contained in a cylinder of diameter $d_{min}$ and of length $d_{min} \pi \sqrt{2}$ corresponding to the most amplified wavelength of the RP instability \cite{Jens_rmp97}. Hence, the diameter of the emitted droplets $a = (3 \pi \sqrt{2} / 2)^{1/3} \times d_{min} \simeq 1.88 \times d_{min}$. This is indeed close to what is experimentally measured, as $a$ is found to be equal to 1.65 to 1.99 times $d_{min}$. Consequently, we take $d=d_{min}$ in the definition of We.

\begin{figure}[tb]
\begin{center}
\includegraphics[scale=0.57]{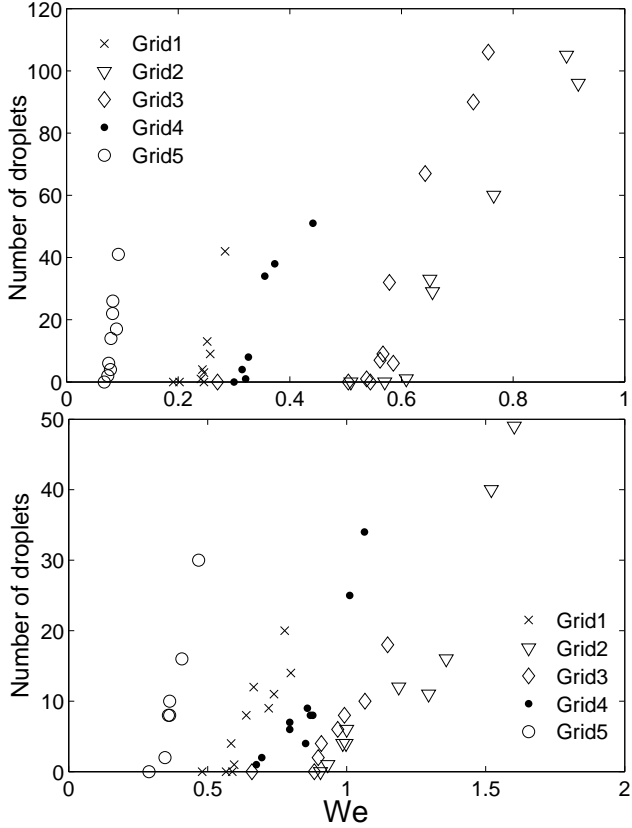}
\caption{Number of emerging droplets for direct (Top) and inverse (Bottom) impact configurations.}
\label{fig:plot_volume}
\end{center}
\end{figure}

\begin{figure}[b]
\begin{center}
\includegraphics[scale=0.315]{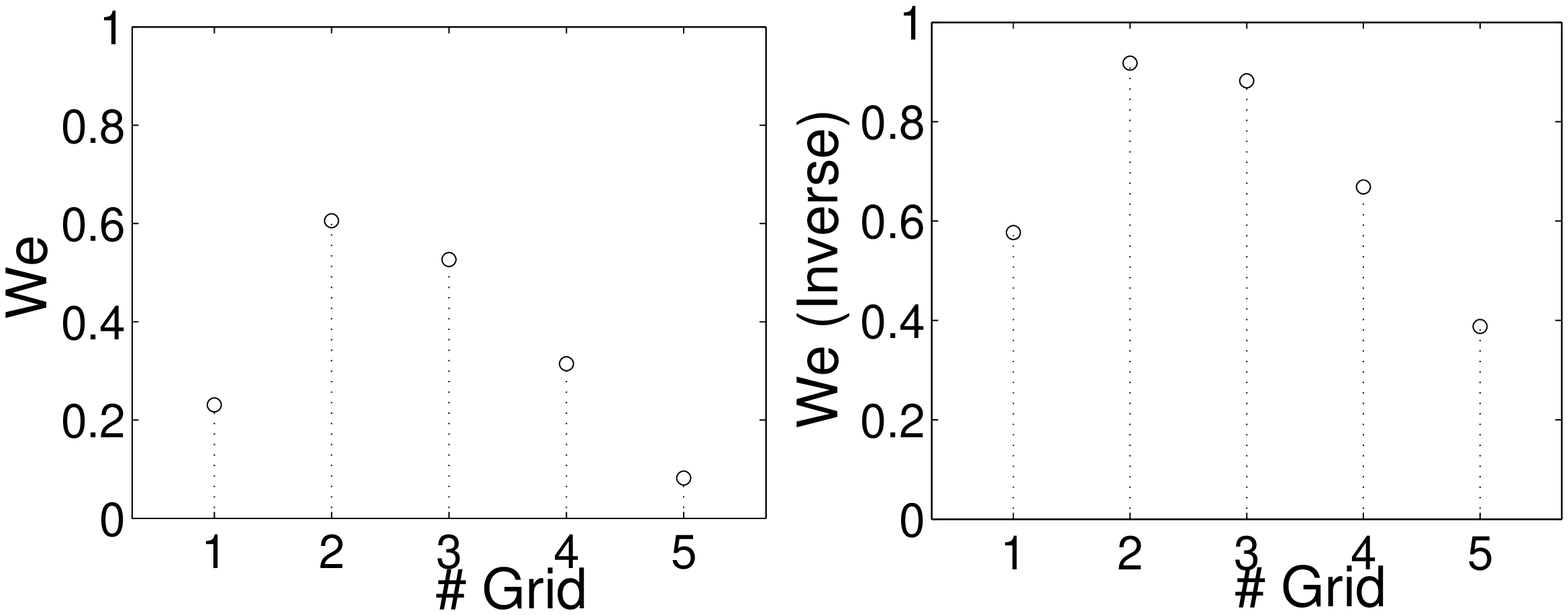}
\caption{Weber number at threshold for direct (\textit{Left}) and inverse (\textit{Right}) impact configurations.}
\label{fig:plot_seuils}
\end{center}
\end{figure}

Figure \ref{fig:plot_volume} shows the number of emerging droplets versus the We, for 5 different grids, and for direct (a) and inverse (b) configurations. The general trend is an increase in droplets number at larger We. An increase of initial kinetic energy leads to a larger final capillary surface energy, hence to a larger number of droplets. We also studied the threshold We for which a single droplet emerges from the grid (see Fig.~\ref{fig:mediumgrid_seq}-(b)). Despite the singular and unpredictable dynamics generally observed at impact \cite{Yarin06,Nonlinearity}, the threshold is well reproducible. The values of threshold We for the five grids are plotted in Fig.~\ref{fig:plot_seuils}. 

\begin{figure}[tb]
\begin{center}
\includegraphics[scale=0.42]{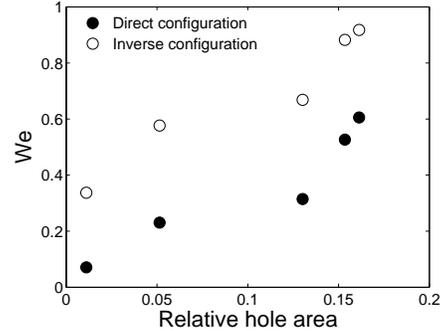}
\caption{Threshold We versus relative area of the holes.}
\label{fig:plot_1mwe}
\end{center}
\end{figure}

It is noteworthy that the threshold We is smaller when $d_{min}$ is small, and it is much smaller than those measured for a single hole (a value close to 3.9 was found in \cite{Lorenceau03}). These results suggest that there are collective effects leading to an additional pressure that contributes to make the liquid penetrate more easily through the grid. As for impacts on textured surfaces \cite{Dengetal09}, this pressure is due to the shock occurring when the liquid experiences a sudden change of its momentum. In the present case, the relative contribution of this additional pressure is larger for smaller holes. Hence, it is natural to interpret this effect as a sort of \textit{water hammer pressure}: this effect is more prominent if more liquid is forced to change direction, hence if the relative area of the holes on the impact surface is small. This is what is shown in Fig.~\ref{fig:plot_1mwe}. The threshold We, decreasing when the relative contribution of the additionnal pressure is larger, is plotted versus the relative area of the holes $d_{min}^2 / d_h^2$ ($d_h$ is the distance between two holes centers). The water hammer pressure is of the same order as the dynamical pressure $P_{dyn} = 1/2 \rho U^2$ and the capillary pressure $P_{cap} = \sigma / d_{min}$, but it is not necessary homogeneously distributed on the impact surface. This leads to surprising effects like that depicted in Fig.~\ref{fig:mediumgrid_seq}-(c): there is a remarkable absence of droplets at the centre of impact. Most droplets emerge away from the centre, on the contrary to Fig.~\ref{fig:mediumgrid_seq}-(a) where more droplets emerge (with a larger velocity) at the centre. The latter case corresponds to the most porous grid. The double-peak is reminiscent to what is recently predicted by Mandre \textit{et al.} \cite{Mandre_09} for an impact on a solid surface. Figure \ref{fig:plot_seuils} also reveals that the threshold We is \textit{always larger} in the inverse configuration than in the direct one, for a given grid: this is to be related to recent observations and explanation that texture with re-entrant shape - like in our grids in inverse configurations - provides a good protection against liquid impalement \cite{Tuteja08}.

In summary, our experiments showed an original way to produce a monodisperse spray of droplets, which size is governed by the smallest size of an individual hole. This simple system is useful to understand the impalement of drops on textured surfaces \cite{Bartolo06}. For droplet manipulations on frictionless surfaces, our grids constitute bottomless substrates, which can be an asset for reverting the impalement transition (Wenzel-to-Cassie state \cite{Cassie_Baxter_Wenzel}), something that is not possible with usual textured surfaces.

\end{document}